\documentstyle[prl,aps,multicol,epsfig]{revtex}
\begin{document} 
\draft 
\title{Anomalous low doping phase of the Hubbard model}
\author{C.\ Gr\"ober, R.\ Eder and W.\ Hanke}
\address{Institut f\"ur Theoretische Physik, Universit\"at W\"urzburg,
Am Hubland,  97074 W\"urzburg, Germany}
\date{\today}
\maketitle
\begin{abstract}
We present results of a systematic Quantum-Monte-Carlo study for the single-band Hubbard model.
Thereby we evaluated single-particle spectra (PES \& IPES), two-particle spectra (spin \& density
correlation functions),
and the dynamical correlation function of suitably defined diagnostic operators, all
as a function of temperature and hole doping. The results allow to identify different physical 
regimes. Near half-filling we find an anomalous  `Hubbard-I phase', where the band structure is, 
up to some minor modifications, consistent with the Hubbard-I predictions. At lower
temperatures, where the spin response becomes sharp, additional dispersionless
`bands' emerge due to the dressing of electrons/holes with spin excitatons. We present a simple 
phenomenological
fit which reproduces the band structure of the insulator quantitatively. 
The Fermi surface volume in the low doping phase,
as derived from the single-particle spectral function, is not consistent with the Luttinger theorem,
but qualitatively in agreement with the predictions of the Hubbard-I approximation.
The anomalous phase extends up to a hole concentration of $\approx 15$\%, i.e. the underdoped
region in the phase diagram of high-T$_c$ superconductors.
We also investigate the nature of the magnetic ordering transition in the single particle spectra.
We show that the transition to an SDW-like band structure is not accomplished by the
formation of any resolvable `precursor bands', but rather by a (spectroscopically invisible)
band of spin 3/2 quasiparticles. We discuss implications for the
`remnant Fermi surface' in insulating cuprate compounds and the shadow bands in the
doped materials.
\end{abstract} 

\begin{multicols}{2}

\section{Introduction}

The 1-band Hubbard model on a two-dimensional square lattice has the Hamiltonian 
\begin{eqnarray}
  H &=& -t \sum_{\langle i,j \rangle,\sigma} (c^{\dagger}_{i,\sigma} c^{}_{j,\sigma} + h.c.)
  + U \sum_{i} (n_{i,\uparrow}-\frac{1}{2}) (n_{i,\downarrow}-\frac{1}{2}) \nonumber \\
  && - \mu \sum_{i} (n_{i,\uparrow} + n_{i,\downarrow}).
  \label{1bandhubbard}
\end{eqnarray}
Here, $c^{\dagger}_{i,\sigma}$ ($c^{}_{i,\sigma}$) creates (annihilates) an electron with spin
$\sigma$ in a Wannier orbital centered at lattice site $i$. The particle density at each site 
is given by $n_{i,\sigma}=c^{\dagger}_{i,\sigma} c^{}_{i,\sigma}$. The first sum for the kinetic 
energy is restricted to include only the hopping matrix element $t$ between next-nearest neighbor 
sites $\langle i,j \rangle$. Periodic boundary conditions are used throughout the following work. 
The second sum describes for $U>0$ an on-site Coulomb repulsion between particles of opposite spin 
that share the same lattice site. In the present work we restrict ourselves to $U=8.0 \, t$. The 
chemical potential $\mu$ in the third sum controls the occupation of the finite lattice in the 
finite-temperature grand canonical Quantum-Monte-Carlo (QMC) simulation we preformed. 
At half-filling, particle-hole 
symmetry of the kinetic and $U$-term implies $\mu=0$. The analytic continuation of the dynamic 
imaginary-times QMC data to the real frequency axis is performed with state-of-the-art Maximum-Entropy 
(ME) techniques. For exhausting discussions concerning the QMC and ME methods we refer the reader 
to \cite{Loh92,Hirsch85,Jarrel96}.\\
The 1-band Hubbard model exhibits several energy scales: in the repulsive case the high-energy 
scale $U$ is important in determining the insulating gap at half-filling, $\langle n \rangle=1.0$. 
An important low-energy scale is set by the exchange interaction $J=4t^{2}/U$: in second order 
perturbation theory two particles with different spins on neighboring lattice sites can exchange 
via a virtual double occupation. This process is the source for the strong antiferromagnetic (AF) 
correlations found near and at half-filling.\\
With the exception of some known symmetry properties like invariance under global spin-rotation, 
whose generators form the $SU(2)$--algebra, and invariance under $U(1)$-transformation, i.e., 
charge-conservation, as well as the particle-hole transformation, no rigorous results are known 
for the 1-band Hubbard model in two-dimensions \cite{Fradkin91}. The Mermin-Wagner theorem 
\cite{Mermin66} prevents a long-range ordered state in a two-dimensional system for finite 
temperatures, but it is commonly believed that the ground state of the spin-$1/2$ Heisenberg
antiferromagnet, i.e. the large-$U$ limit of the repulsive half-filled Hubbard 
model \cite{Harris67}, shows
long-range N\'eel order in two dimensions.
N\'eel order results also in the weak-coupling limit 
\cite{Schrieffer88,Schrieffer89}, 
where the gap $\Delta$ is due to a spin-density-wave (SDW) instability related 
to perfect nesting.\\
It might seem that due to the Mermin-Wagner theorem the physics of the ordered phase
is out of reach for our numerical technique, which is limited to
finite temperatures.
However, one may assume that the system `effectively orders'
as soon as the spin-correlation length becomes comparable to the system size.
Due to the periodic boundary conditions
the spin-correlation function
$\chi(\bbox{r})=(1/L^2)\sum_i \langle \vec{S}_i \cdot \vec{S}_{i+\bbox{r}} \rangle$ is a periodic
function of $\bbox{r}$ and if the value of this function at the
maximum value $|\bbox{r}|=\sqrt{2L}$ (with $L$ $\equiv$ cluster size) 
is still appreciable, we may expect that the system is `effectively ordered'. 
A rough measure would be the spin-correlation length
$\zeta$, obtained by fitting $\chi(\bbox{r})$ to the form
$a \cdot \vert {\bf r} \vert^{b} \cdot e^{-\vert {\bf r} \vert/\zeta}$.
Since the infinite system has the N\'eel temperature
$T=0$, the spin-correlation length
diverges as $T \rightarrow 0$ and we may expect that in a {\em finite} system
$\zeta$ becomes comparable to the cluster size $L$ at a {\em finite}
temperature which depends on the lattice size. Below this temperature we then expect that
the system resembles the ordered phase, although the spin-rotation symmetry
persists even in this case.
In that sense, the finite size of the system creates an artificial
`N\'eel temperature' which, however, depends on $L$ and $U/t$ etc and has no
real counterpart in the infinite system.
This has to be kept in mind when discussing the results.\\
We proceed by discussing some known approximations to the Hubbard model.
The `classical' approximation to the Hubbard model is the so-called Hubbard-I 
approximation\cite{hubbard}. Its essence is the
splitting of the electron annihilation operator into the two 
`eigenoperators' of the interaction part: 
\begin{eqnarray}
  c^{}_{i,\sigma} &=& c^{}_{i,\sigma} n_{i,\bar \sigma} + c^{}_{i,\sigma} 
  (1-n_{i,\bar \sigma}) \nonumber \\
  &=& \frac{1}{\sqrt{2}}(d^{}_{i,\sigma} + h^{\dagger}_{i,\sigma}), 
\label{interactioneigenoperators}
\end{eqnarray}
whence
\begin{eqnarray}
  \lbrack d^{}_{i,\sigma}, H_{U} \rbrack &=& \frac{U}{2} d^{}_{i,\sigma}, \nonumber \\
  \lbrack h^{\dagger}_{i,\sigma}, H_{U} \rbrack &=&-\frac{U}{2}h^{\dagger}_{i,\sigma}.
 \label{commuteinteraction}
\end{eqnarray}
The physical content of the Hubbard-I approximation, 
which neglects any spin-correlations,
becomes clear by realizing \cite{ansgar} that the
equations of motion in this approximation are completely equivalent to
an `effective Hamiltonian' for double occupancy-like particles 
$d^{\dagger}_{i,\sigma}=(1/\sqrt{2}) \, \, \, c^{\dagger}_{i,\sigma}n_{i,\bar \sigma}$ 
and hole-like particles $h^{\dagger}_{i,\sigma}=(1/\sqrt{2}) \, \, \, c^{}_{i,\sigma}
(1-n_{i,\bar \sigma})$:
\begin{eqnarray}
H &=& -\frac{t}{2} \sum_{\langle i, j \rangle,\sigma} (
d_{i,\sigma}^\dagger d_{j,\sigma} 
- h_{i,\sigma}^\dagger h_{j,\sigma} )
\nonumber \\
&& -\frac{t}{2} \sum_{\langle i, j \rangle,\sigma} (
d_{i,\sigma}^\dagger h_{j,\bar{\sigma}}^\dagger + h.c.)
\nonumber \\
&+& \frac{U}{2} \sum_{i,\sigma} (d_{i,\sigma}^\dagger d_{i,\sigma} 
- h_{i,\sigma}^\dagger h_{i,\sigma} ).
\label{hubeff}
\end{eqnarray}
This Hamiltonian contains terms which describe the pair creation of a hole and
a double occupancy on nearest neighbors $\langle i,j \rangle$, terms which 
describe the propagation of these effective particles, and an additional energy 
of formation of $U$ for the double occupancy. The matrix elements for the 
propagation are reduced by a factor of $1/2$, because in an uncorrelated
spin-background there is a probability of $1/2$ for the spin on a nearest neighbor
to have the proper direction to allow for the hopping of an electron.
Solving (\ref{hubeff}) by Fourier and Bogoliubov transform:
\begin{eqnarray}
\gamma_{1,\bbox{k},\sigma} &=& u_{\bbox{k}} d_{\bbox{k},\sigma}
+ v_{\bbox{k}} h_{-\bbox{k},\bar{\sigma}}^\dagger,
\nonumber \\
\gamma_{2,\bbox{k},\sigma} &=&-v_{\bbox{k}} d_{\bbox{k},\sigma}
+ u_{\bbox{k}} h_{-\bbox{k},\bar{\sigma}}^\dagger,
\end{eqnarray}
yields the standard
dispersion relation,  which consists of the upper and lower Hubbard bands:
\begin{equation}
  E^{Hub-I}_{\pm} ({\bf k})=\frac{1}{2} \left( \epsilon_{\bbox{k}}
    \pm \sqrt{\epsilon_{\bbox{k}}^{2} + U^{2}} \right). \label{hubIhafdisp}
\end{equation}
Here $\epsilon_{\bbox{k}}$ denotes the free tight-binding dispersion 
$\epsilon_{\bbox{k}}=-2t(\cos(k_{x})+\cos(k_{y}))$. 
Using (\ref{interactioneigenoperators})
we also obtain the correct spectral weights of the two Hubbard bands:
\begin{eqnarray}
  Z^{Hub-I}_{\pm} ({\bf k}) &=&
\frac{1}{2}(u_{\bbox{k}} \pm v_{\bbox{k}})^2
\nonumber \\
&=&
\frac{1}{2} \left(1 \pm \frac{\epsilon_{\bbox{k}}}{\sqrt{\epsilon_{\bbox{k}}^2 + U^2} } \right).
\label{hubIhafwght}
\end{eqnarray}
As noted above, the key assumption in the Hubbard I approximation is the neglect of
spin-correlations. Therefore, we expect that this approximation will become inaccurate
as soon
as spin-correlations become sufficiently strong so as to
appreciably influence the propagation of holes and double occupancies. This effect
will be strongest
in the N\'eel ordered phase believed to be realized in the ground state.
If we choose the `spin-background', in which the
holes and double occupancies propagate, to be the N\'eel state, the double occupancy
$d_{i\uparrow}^\dagger$ with spin $\uparrow$ can exist only on the 
$\downarrow$ sublattice and vice
versa. Similarly, a hole $h_{i\downarrow}^\dagger$ can be created only
on the $\downarrow$ sublattice and vice versa. We, thus, expect that we
have to modifiy the Hubbard-I Hamiltonian (\ref{hubeff}) into:
\begin{eqnarray}
H &=& 
 -t \sum_{ i\in A, j\in N(i)} (
d_{i,\uparrow}^\dagger h_{j,\uparrow}^\dagger + h.c.)
\nonumber \\
&& -t \sum_{ i\in B, j\in N(i)} (
d_{i,\downarrow}^\dagger h_{j,\downarrow}^\dagger + h.c.)
\nonumber \\
&+& \frac{U}{2} \sum_{i,\sigma}( d_{i,\sigma}^\dagger d_{i,\sigma}
 - h_{i,\sigma}^\dagger h_{i,\sigma}),
\label{swham}
\end{eqnarray}
where $A$ ($B$) denotes the $\downarrow$ ($\uparrow$) sublattice and 
$N(i)$ denotes the set of nearest neighbors of site $i$.
Note that the $d_{i,\sigma}^\dagger d_{j,\sigma}$ and $h_{i,\sigma}^\dagger h_{j,\sigma}$
propagation terms drop out (because of the $A$ and $B$ sublattices), 
and the matrix element for pair creation are $t$ rather than $t/2$ - this
takes into account the fact that the spins on neighboring sites are
antiparallel with probability $1$ rather than $1/2$ as was the case 
in the paramagnetic state.
Fourier and Bogoliubov transformation of (\ref{swham}) yields the dispersion
\begin{equation}
E_{\pm}^{SDW}(\bbox{k}) = 
\pm \sqrt{ \epsilon_{\bbox{k}}^2 + \frac{U^2}{4} },
\end{equation}
and using
\begin{eqnarray}
c_{\bbox{k},\uparrow} &=& h_{\bbox{k},\uparrow}
 + d_{-\bbox{k},\downarrow}^\dagger
\nonumber \\
c_{\bbox{k}+\bbox{Q},\uparrow} &=& h_{\bbox{k},\uparrow}
 - d_{-\bbox{k},\downarrow}^\dagger
\end{eqnarray}
(where $\bbox{k}$ is within the AF Brillouin zone),
we find the spectral weight of these bands:
\begin{eqnarray}
Z_{\pm} &=& \frac{1}{2}( u_{\bbox{k}} \pm v_{\bbox{k}})^2
\nonumber \\
&=& \frac{1}{2}\left( 1 \pm \frac{\epsilon_{\bbox{k}}}
{\sqrt{\epsilon_{\bbox{k}}^2 + \frac{U^2}{4}}}\right).
\label{crep}
\end{eqnarray}
This is precisely what is obtained from the SDW mean-field treatment of the 
Hubbard model by setting the staggered magnetization $m$ to a value of $1$
(which is a good approximation in the limit of large $U$).
In general, the SDW mean-field approximation models the AF N\'eel state by 
assuming $\langle n_{i,\sigma} \rangle=\frac{1}{2}(1+\sigma m 
e^{i {\bf Q}{\bf R}_{i}})$
with AF nesting vector ${\bf Q}=(\pi,\pi)$ and staggered magnetization $m=e^{i {\bf Q}{\bf R}_{i}}
\langle n_{i,\uparrow}-n_{i,\downarrow} \rangle$\cite{Schrieffer88,Schrieffer89}.
This results in the two-band dispersion
\begin{equation}
  E^{SDW}_{\pm} ({\bf k})=\pm \sqrt{\epsilon_{\bbox{k}}^{2}+\Delta^{2}}, \label{sdwdisp}
\end{equation}
and the spectral weight
\begin{equation}
  Z^{SDW}_{\pm} ({\bf k})=\frac{1}{2} \left( 1 \pm \epsilon_{\bbox{k}}/
    E^{SDW}_{+} ({\bf k}) \right). \label{sdwwght}
\end{equation}
The gap parameter is $\Delta=Um/2$ and the staggered magnetization $m$ is determined 
self-consistently from the following equation:
\begin{equation}
  m=\frac{2}{N} \sum^{occupied}_{{\bf k}} \frac{Um}{\sqrt{(\epsilon_{\bbox{k}}
      -\epsilon({\bf k}+{\bf Q}))^{2} + m^{2} U^{2}}}. \label{sdwgap}
\end{equation}
The solution of this self-consistency equation yields the value $\Delta=3.56 \, t$ 
at $U=8.0 \, t$, the value used in the present work.
If we set $m=1$ on the other hand, as would be
apropriate for $U/t \rightarrow \infty$, we obviously recover the results from our
Hubbard-I-like Hamiltonian (\ref{swham}).\\
In the two limiting cases of no spin-correlations
and of perfect N\'eel order we can thus treat the Hubbard model in
a quite analogous fashion and, as will be seen below, the
Hubbard-I results are indeed a good approximation to the
actual spectral function in the limit of high temperature. The main problem
then is how to describe the effect of spin-fluctuations and 
how to manage the crossover from the completely disordered to the N\'eel 
ordered phase. Below, we will adress this crossover by QMC simulations.\\ 
Thereby we want to take advantage
of the possibility to calculate the spectra
of specifically designed `diagnostic operators'.
The first one of these is the {\it `shadow' operator}:
\begin{eqnarray}
\tilde{c}_{i,\sigma} &=& c_{i,\sigma} n_{i,\bar{\sigma}}
- c_{i,\sigma} (1-n_{i,\bar{\sigma}}) \nonumber \\
&=& \frac{1}{\sqrt{2}}(d^{}_{i,\sigma} - h^{\dagger}_{i,\sigma}). \label{shadoweqt}
\end{eqnarray}
We note first of all, that the Fourier transform of this operator, $\tilde{c}_{\bbox{k},\sigma}$
 has  precisely the same quantum numbers as the ordinary electron operator
$c_{\bbox{k},\sigma}$: momentum $-\bbox{k}$, total spin $1/2$, $z$-spin $\sigma$,
and identical point group symmetry at high symmetry momenta.
It follows that the poles in its dynamical correlation functions
\begin{eqnarray}
  A_{\tilde{c}} (\bbox{k},\omega) &=& \sum_\nu |\langle \Psi_\nu|\tilde{c}_{\bbox{k},\sigma} 
  \vert\Psi_{\mu}\rangle\vert^2 
\nonumber \\ 
&& e^{-\beta(E_\mu-E_\nu)}\delta(\omega -( E_\nu - E_0))
\end{eqnarray}
originate from exactly the same final states $| \Psi_\nu \rangle$
as those of the photoemission spectrum. It can happen only
accidentally that a given state $\nu$ has an exactly vanishing
weight in one of the correlation functions, but not the others.
In this case, however, any arbitrarily small perturbation
will remove the accidental vanishing of the peak.
The only thing that can and will be different in the
spectra of the diagnostic operator, are the {\em weights}
of the peaks, $|\langle \Psi_\nu|\tilde{c}_{\bbox{k},\sigma}  |\Psi_0\rangle|^2$.
In fact, comparison of equations (\ref{crep}) and (\ref{hubIhafwght}) shows
immediately that
both for the Hubbard I approximation and for the SDW-approximation
use of the operator $\tilde{c}_{\bbox{k},\sigma}$ instead
of $c_{\bbox{k},\sigma}$ exchanges the
weights of the two Hubbard/SDW bands.
It follows that band portions which have a small spectral weight
in the spectra of the ordinary electron operator
will aquire a large spectral weight in the spectrum of
$\tilde{c}_{\bbox{k},\sigma}$ and vice versa. Therefore, this diagnostic operator is 
a useful tool to map out the `shadowy' parts of the spectra in the outer
parts of the Brillouin zone. An additional benefit
is that since the 
ME technique resolves peaks with large spectral weight
more reliably than those with small weight, we can get
more precise information about the dispersion of these
bands with weak intensity.
The second diagnostic operator that we will be using is
the {\it `spin-$1/2$ string' operator}
\begin{eqnarray}
b_{i,\uparrow}^{} &=& \sum_{j \in N(i)}
(\;c_{i\uparrow}^{} S_{j}^z + c_{i\downarrow}^{} S_j^-\;), \nonumber \\
b^{\dagger}_{i,\uparrow} &=& \sum_{j \in N(i)}
(\;c_{i\uparrow}^\dagger S_{j}^z + c_{i\downarrow}^\dagger S_j^+\;),
\label{stringoperator}
\end{eqnarray}
where the sum on the r.h.s. runs over the nearest neighbors $j$ of site $i$.
This operator is the Clebsch-Gordan contraction of the adjoint rank-$1$ spinor 
$c_{i,\sigma}$ and the spin-$1$ vector operator $\vec{S}_j$ into yet another 
adjoint rank-$1$ spinor. It describes the annihilation of a `dressed' electron, 
i.e., an electron with a spin-excitation on a nearest neighbor. 
Again, since the Fourier transforms of the diagnostic operator,
$b_{\bbox{k},\sigma}$, agrees with the
electron annihilation operator $c_{\bbox{k},\sigma}$ in all
conceivable quantum numbers (momentum, total spin, $z$-component
of the spin, point group operations in the case of
high symmetry momenta), it obeys precisely the same selection
rules, whence it is just as good as the
$c$-operator itself to map out the band structure.\\
One further point, which is important from the
technical point of view, is the following:
under the particle-hole transformation $c_{i,\sigma} \rightarrow
e^{i \bbox{Q} \cdot \bbox{R}_i} c_{i,\sigma}^\dagger$ we have
$\tilde{c}_{i,\uparrow}^{} \rightarrow -e^{i \bbox{Q} \cdot \bbox{R}_i}
\tilde{c}_{i,\uparrow}^\dagger$
and $b_{i,\uparrow}^{} \rightarrow -e^{i \bbox{Q} \cdot \bbox{R}_i}
b_{i,\uparrow}^\dagger$.
This implies
that at half-filling the spectra of the shadow and spin-$1/2$ string operator obey the 
same particle-hole-symmetry as those of the ordinary electron operator, i.e. 
\begin{equation}
A(\bbox{k},\omega) =
A(\bbox{k}+\bbox{Q}, -\omega).
\label{phsym}
\end{equation}
In our MaxEnt program
particle-hole-symmetry is not implemented as an additional constraint, 
in other words: the MaxEnt procedure does not `know' about this 
additional symmetry. The degree to which (\ref{phsym}) is obeyed
in the final spectra thus gives a good check
for the accuracy of the reconstructed spectra. This is of particular
importance in the case of the spin-$1/2$ string operator because the
Wick-contraction of this operator on any given time slice produces a 
total of $\approx 80$ products of noninteracting Green's functions.
The computation is, therefore, much more prone 
to inaccuracies so that an additional check is 
 desirable.\\
The present work is organized as follows: first, we compare the temperature 
dependent dynamic single-particle properties of the Hubbard model with 
the predicitions of the mean-field SDW and Hubbard-I approximations. In addition,
we consider the temperature dependent two-particle excitations. 
Then, we will use our first diagnostic operator, the shadow operator 
$\tilde{c}_{\bbox{k},\sigma}$, to show the existence of a total of four 
bands in the photoemission spectrum and to shed light onto the temperature 
dependent crossover from the SDW to the Hubbard-I regime. Then we will 
investigate the 4-band structure in more detail: after a phenomenological 
fit, which is able to produce a total of four bands, we will consider the 
string picture which naturally leads to our second diagnostic operator, 
the spin-$1/2$ string operator $b_{\bbox{k},\sigma}$. This operator will 
be used to ultimately reveal the underlying mechanisms behind the generation
of the 4-band structure, namely the dressing of the photoholes by clouds of 
AF spin-excitations.
To resolve the `AF mirror image' of the narrow quasiparticle spectral weight 
features between ${\bf k}=(0,0)$ and ${\bf k}=(\pi,0)$ around momentum 
${\bf k}=(\pi,\pi)$ we introduce also a spin-$3/2$ string operator.
Finally, we will concentrate on the doped regime, thereby showing
the violation of the Luttinger theorem near half-filling, and discuss the 
hole concentration range 
in which these dressing effects dominante the low-energy physics.
\section{Temperature dependent dynamics at half-filling}
We start in figure \ref{fig1} with the discussion of the angle-resolved single-particle 
spectral function $A({\bf k},\omega)$ for $U=8.0 \, t$ and various temperatures 
in the range from $T=4.00 \, t$ to $T=0.10 \, t$. In the left column of the figure the spectral 
functions are shown as `grey-scale' plots versus momentum ${\bf k}$ and energy $\omega/t$
with dark (light) areas corresponding to large (small) spectral 
weight. The same spectra are shown also in the right column, but now as line-plots at each 
momentum ${\bf k}$. The QMC data in the figure are compared to the renormalized results of the 
Hubbard-I approximation at high and medium temperatures and with the renormalized results of the 
mean-field SDW approximation at the lowest temperature, $T=0.10 \, t$. 
`Renormalized' means here that we have readjusted the parameters
$U$ and $t$ in (\ref{hubIhafdisp}) and
(\ref{sdwdisp})so as to obtain an optimal fit to the `bands' of high spectral weight
in the spectra. These approximate  dispersions 
are plotted as solid lines in the left column, while their spectra are 
shown in the right column as line-plots at each momentum ${\bf k}$.
Thereby we have assumed a Lorentzian lineshape with 
a suitably chosen temperature dependent width (the Hubbard-I approximation does not
provide any information about linewidths).\\
Starting at the highest calculated temperature, $T=4.00 \, t$, we find that the 
Hubbard-I approximation 
with practically unrenormalized parameters 
($\tilde t=0.95 \, t$ and $\tilde U=8.32 \, t$) 
fits the QMC spectral functions almost 
perfectly regarding both the 
general dispersion and the distribution of spectral weight
(quadratic deviation per degree of freedom $\chi^{2}=0.05$). 
This is not surprising since the 
Hubbard-I approximation is derived for the paramagnetic state 
thereby neglecting all effects 
of spin-correlations, i.e., the Hubbard-I approximation essentially describes 
the interplay between itinerant 
electrons and strong on-site repulsion. This should be a reasonable assumption at 
this high temperature since 
all relevant spin-degrees of freedom are thermally excited.\\
Lowering the temperature to $T=1.00 \, t$ and further to $T=0.33 \, t$, the data show that the 
Hubbard-I approximation increasingly fails to
reproduce the entire spectrum and
is only able to fit the peaks with maximal spectral weight reasonably well.
Even then, in order to achieve these fits, one already has to renormalize the free parameters 
strongly. 
The values we found are $\tilde t=1.38 \, t$ and $\tilde U=5.57 \, t$ with $\chi^{2}=1.54$ at 
$T=1.00 \, t$ and $\tilde t=1.40 \, t$ and $\tilde U=3.96 \, t$ with $\chi^{2}=0.85$ 
at  $T=0.33 \, t$. The peaks that are missed by the Hubbard-I approximation are the
states that form the first ionization/affinity states around momentum ${\bf k}=(0,0)/(\pi,\pi)$ on 
the photoemission/inverse photoemission side and two rather dispersionless bands at 
higher energies of $\omega \approx \pm 6.0 \, t$. 
The former states were previously resolved by Preuss 
and co-workers \cite{Preuss95}. Alltogether one can distinguish
a total of four `bands' in the single particle spectral density.
As will be seen below, the temperature/doping regime where this 4-band 
structure is seen coincides with the regime where a
collective low-energy mode with momentum $(\pi,\pi)$ in the spin-response exists.
Inspite of this, however,  we stress 
that the 4-band structure cannot be explained by a backfolding 
of the band structure due to ordering effects 
since the spin-correlation length is $\leq 1.5$ lattice spacings 
at temperatures $T \geq 0.33 \, t$.\\ 
For the lowest temperature, $T=0.10 \, t$, the QMC data are compared with the results from the 
AF SDW approximation. As in the case of the Hubbard-I approximation at medium temperatures, the 
lowermost spectra of figure \ref{fig1} show that the SDW approximation is only able 
to fit the peaks with large spectral weight. Again one has to renormalize the 
free parameters heavily to values of $\tilde t=1.34 \, t$ and $\tilde \Delta=2.29 \, t$ with 
$\chi^{2}=0.83$. Moreover, as was the case for the Hubbard-I approximation
at higher temperatures, the SDW approximation neither 
explains the states that form the first ionization/affinity states around momentum 
${\bf k}=(0,0)/(\pi,\pi)$ on the photoemission/inverse photoemission side, nor the 
two dispersionless bands at higher excitation energies which can be seen rather clearly
in the spectra.\\
All in all, the overall distribution of spectral weight is 
roughly reproduced by the Hubbard-I and SDW 
approximations as long as one forgets about the 4-band structure.
In fact it is well known that the {\em integrated} 
photoemission or inverse photoemission weight 
(that means the electron momentum distribution) at each ${\bf k}$--point 
is reproduced quite well by the Hubbard-I approximation and the related
$2$-pole approximation\cite{bernhard}. \\
As already mentioned,
the emerging of the 4-band structure in the photoemission somewhere in between $T=1.00 \, t$ 
and $T=0.33 \, t$ is closely related to a change in the spin-response: 
to illustrate this we consider figure \ref{fig2}, which shows 
the spin-correlation function, $\chi_{sz}({\bf k},\omega)$ (left column), and the 
charge-correlation function, $\chi_{cc}({\bf k},\omega)$ (right column), for 
different temperatures.
Whereas the spin-response is entirely incoherent at $T=1.00 \, t$,
with decreasing temperature
it can be fitted increasingly well by the spin-wave dispersion
\begin{equation}
E^{SW}({\bf k})=2J\sqrt{1-\frac{1}{4}(\cos(k_{x})+\cos(k_{y}))^{2}}. \label{spinwave}
\end{equation}
This result is known from previous calculations \cite{Schrieffer88,Schrieffer89}, which 
demonstrated that the two-particle correlation functions like the spin-response 
can be described 
within the SDW approximation even for large values of the interaction $U$. 
The energy scale $J$ directly manifests itself
in the spin-response since the spin-wave dispersion takes the value of $E^{SW}(\pi,0)=2J$ at 
momentum ${\bf k}=(\pi,0)$. The fit parameters are $\tilde J=0.33 \,t$ with $\chi^{2}=0.01$ at 
 $T=0.33 \, t$ and $\tilde J=0.49 \, t$ with $\chi^{2}=0.11$ at  $T=0.10 \, 
t$. The latter is already quite close to the strong coupling
estimate $J=4t^2/U=0.5t$. 
Furthermore, the figure shows that with decreasing 
temperature the spin-response concentrates its weight 
more and more at the AF momentum ${\bf Q}=(\pi,\pi)$ (as it is the case 
in AF spin-wave theory) and at a characteristic energy $\omega^{\star}$. 
The latter decreases with 
decreasing temperature, i.e., the spin-response comes closer and closer to the predictions 
of AF spin-wave theory (\ref{spinwave}). 
The spin-correlation length $\zeta_{sz} (T)$ can be derived from a real-space fit of the QMC 
equal-imaginary-times spin-correlation function $\chi(\bbox{r})$
to the form $a \cdot \vert {\bf r} \vert^{b} \cdot 
e^{-\vert {\bf r} \vert/\zeta_{sz} (T)}$ thereby
incorporating the effects of the periodic boundary conditions. 
While this is the best one can do on a finite lattice with periodic boundary conditions, the fit will 
only lead to roughly correct values due to the relative small system size of $8 \times 8$. The 
values obtained for the spin-correlation length then
are $\zeta_{sz}=0.3-0.5$, $\zeta_{sz}=1.0-1.3$, $\zeta_{sz}=
1.6-1.9$, $\zeta_{sz}=2.1-2.8$ and $\zeta_{sz}>8$ for $T=1.00 \, t$, $T=0.33 \, t$, $T=0.25 \, 
t$, $T=0.20 \, t$ and $T=0.10 \, t$, respectively. We actually believe that the spin-correlation length 
reaches the system size already at a temperature of $T \approx 0.20 \, t$, because at this temperature the 
fit results in values between $\zeta_{sz}=2.1$ and $\zeta_{sz}=2.8$ (with the exponent $b$ set to zero or 
not) but always with error bars of roughly the system size. \\
The charge-response $\chi_{cc}({\bf k},\omega)$, on the other hand, is 
rather broad in both momentum ${\bf k}$ and energy $\omega$ for all temperatures studied. 
Furthermore, the charge-response is  gapped for temperatures below $T \approx 1.00 \, t$ 
and, therefore, can certainly not be responsible for any low-energy features of 
the single particle spectrum.\\
It is then quite obvious that at roughly the 
same temperature where the two narrow dispersive 
quasiparticle-like bands (that cannot be 
interpreted within the framework of the Hubbard-I or SDW approximations) 
appear in the single particle 
spectrum, the spin-response develops a sharp collective low-energy mode. We 
conclude that the underlying mechanism behind the occurrence of the 4-band structure 
consists in dynamical magnetic correlation effects, which are beyond the scope of the 
Hubbard-I and SDW approximations. \\
In the following, we want to explore the single particle spectrum
by means of our {\it diagnostic operators}. The first of them is the shadow operator 
$\tilde{c}_{i,\sigma}$ of equation (\ref{shadoweqt}) which will be used to transfer spectral 
weight from the inner parts of the Brillouin zone to the outer ones in case of normal photoemission 
$\omega < \mu$. As already discussed
above, this also improves the resolution of the ME method in this 
region, since its resolution strongly depends on the spectral weight at a certain position. 
Nevertheless
the spectrum of the shadow operator has to exhibit exactly the same peak positions as 
the normal photoemission spectrum.\\
Figure \ref{fig3} shows the angle-resolved spectral function 
of the shadow operator for moderate and low temperatures.
As expected, the shadow operator has its main 
spectral weight near ${\bf k}=(\pi,\pi)$ on the photoemission side and near
${\bf k}=(0,0)$ on the inverse photoemission side. Furthermore, it's spectrum
supports the existence of a total of four bands, because it 
resolves a group of peaks forming dispersionless bands at energies of $\omega \approx \pm 6.0 
\, t$, a region where the normal photoemission spectrum exhibits only some weak and smeared-out
spectral weight. 
These two dispersionless bands at energies of $\omega \approx \pm 6.0 \, t$ are
inconsistent with the dispersions of the Hubbard-I and SDW approximations 
of figure \ref{fig1}. We will further address this topic later in this 
work.\\
Next, we turn in more detail to the temperature dependence of the photoemission spectrum. Figure 
\ref{fig4} shows some close-ups of the normal photoemission spectrum and of the spectrum of 
the shadow operator at momentum ${\bf k}=(\pi,\pi)$ and different temperatures.
For the normal photoemission operator these close-ups show
a peak at $\omega \approx -1.5 \, t$, which would be
consistent with Hubbard-I (see Figure \ref{fig2}). 
In the spectrum of the shadow operator this feature is visible as 
a single resolved peak only at the highest temperature, 
$T=1.00 \, t$ whereas for temperatures down to $T=0.20 \, t$ there is only some diffuse weight
at this position. 
In the ordinary photoemisson spectrum the peak looses spectral weight
with decreasing temperature. It disappears completely at
$T<0.20 \, t$ where the spin-correlation length $\zeta_{sz} (T)$ reaches the 
system size (see above). 
Thus, the temperature $T \approx 0.20 \, t$ 
where we lose the Hubbard-I-like peak at $\omega \approx -1.5 \, t$ 
and ${\bf k}=(\pi,\pi)$ 
coincides quite accurately with the temperature where `effective' long-range order sets in.
Moreover, we find that
as the normal photoemission spectrum loses the peak at $\omega \approx -1.5 \, t$ and
${\bf k}=(\pi,\pi)$, the spectrum 
of the shadow operator gains weight at $\omega \approx -6.0 \, t$. Thus, we 
expect that both features are closely related to the temperature development of the spin-correlation 
length $\zeta_{sz} (T)$.
We note, however, that the crossover in the shape of the dispersion
from Hubbard-I-like to SDW-like occurs in a quite unexpected way:
the topmost band at $(\pi,\pi)$ does {\em not} deform into the SDW form in
any continuous way, but simply `fades away' and eventually
vanishes at the transition.\\
A further surprising result is the following:
at $T=0.10t$ neither the ordinary electron operator nor the shadow operator
pick up the `AF umklapp band'
corresponding to the narrow dispersive band seen for example
at $\omega \approx -3 \, t$ for $\bbox{k}=(0,0)$, i.e.
there is no corresponding band at $\omega \approx -3 \, t$ and $\bbox{k}=(\pi,\pi)$.
Note that in the framework if the
SDW-approximation the shadow operator {\em must} reproduce this umklapp-band
at $(\pi,\pi)$ with the {\em same weight} as the original band
at $(0,0)$ in the ordinary photoemission spectrum - 
see the discussion in the first section.
That this is not the case shows that even at this lowest temperature, a simple 
SDW-like description of the band structure is invalid, in that the
band structure cannot be understood by simple backfolding of the
spectrum obtained without broken symmetry.
As we will see in the following the 
AF SDW state provides only the `background' for the dressing 
of the photoholes with AF spin-excitations
which dynamically generate a total of four bands. 
\section{The 4-band structure}
We return to the discrepancy between the 2-band dispersions of the Hubbard-I 
and SDW approximations and the 4-band structure actually observed
for example in the spectrum of the shadow operator of figure \ref{fig3}.\\
In order to generate a `4-band structure' out of the two bands of the Hubbard-I and SDW approximations 
we try as a phenomenological ansatz to mix the dispersions 
of the Hubbard-I/SDW approximation 
with two dispersionless 
bands at energies of $E_\pm=\pm 3.0 \, t$. In other words,
for both the photoemission and inverse photoemission spectrum
we diagonalize an `effective' $2\times 2$ Hamilton matrix:
\begin{equation}
  H_{\pm} = \left( 
    \begin{array}{cc}
      E^{Hub-I/SDW}_{\pm} & V \\
      V & E_{\pm} \\
    \end{array} 
  \right), \label{mixeqt1}
\end{equation}
and plot in figure \ref{fig5}  the four bands obtained in this way
on top of the spectral density obtained from QMC at $T=0.33 \,t$ and $T=0.10 \, t$.
For comparison the figure also shows the original (i.e. unhybridized) Hubbard-I 
bands plus the two phenomenological 
dispersionless bands at energies of $\omega=\pm 3.0 \, t$. The figure 
shows that the overall agreement between the QMC peak positions and the four bands generated 
by diagonalizing $H_{\pm}$ of equation (\ref{mixeqt1}) is surprisingly good, 
particularly so in view of the fact 
that for both, Hubbard-I and SDW approximation, only 
{\em unrenormalized} parameters were used. In particular, the 
{\em self-consistently} determined value for the SDW gap $\Delta$ of $3.56 \, t$ was used at 
$T=0.10 \, t$.
The only `external' parameter in this figure is the mixing matrix element $V$,
which was set to a value of $1.0 \, t$.\\
Thus, we find in contrast to previous works \cite{Dagotto92}, that 
the introduction of the dispersionless bands reproduces the 
sinlge-particle gap and the width of the quasiparticle band 
correctly without any renormalization of parameters. Rather, the narrowing of the
quasiparticle band and the reduction of the Hubbard gap as compared to the unrenormalized
parameters is brought about by introduction of the dispersionless bands.
This naturally raises the question as to their physical origin.
In the present paper we restrict ourselves to a more phenomenological and
`numerics based' approach. A complementary and more mathematical discussion
is given in Ref. \cite{ansgar}, where an equation of motion approach
similar to Hubbard's original work is pursued.\\
We consider the commutator
of the creation operator for hole-like particles, $h^{\dagger}_{i,\sigma}=
c^{}_{i,\sigma} (1-n_{i,\bar \sigma})$, which annihilates a particle only on a singly occupied 
site, with the kinetic energy of the Hubbard model and find\cite{ansgar}:
\begin{eqnarray}
  \lbrack h^{\dagger}_{i,\uparrow}, H_{t} \rbrack &=& -t \sum_{j \in N(i)} \lbrack 
  (1-\frac{\langle n \rangle}{2}) c^{}_{j,\uparrow} + (c^{}_{j,\uparrow} S^{z}_{i} 
  + c^{}_{j,\downarrow} S^{-}_{i}) \nonumber \\
  && - \frac{1}{2} c^{}_{j,\uparrow} (n_{i}-\langle n \rangle) + c^{\dagger}_{j,\downarrow} 
  c^{}_{i,\downarrow} c^{}_{i,\uparrow}. 
\label{commute}
\end{eqnarray}
Keeping only the first term in the square brackets of equation (\ref{commute}) reproduces the
Hubbard-I approximation \cite{hubbard}. The second term in the square 
brackets describes the dressing of the created hole by a spin-excitation
and is closely related to the spin-$1/2$ string operator of equation 
(\ref{stringoperator}). The third term describes in an analogous fashion
the coupling of the hole to a density fluctuations, whereas the fourth term describes
the coupling to the $\eta$-excitation\cite{yang}. The two latter types of excitation are
not important for a large positive $U$ near half-filling, $\langle n \rangle=1.0$, and will be
neglected.
Therefore, the operator $\sum_{j \in N(i)} (c^{}_{j,\uparrow} S^{z}_{i} + c^{}_{j,\downarrow} 
S^{-}_{i})$ is in this case the most important correction over the Hubbard-I 
approximation.
As already stated, it describes a hole dressed by a spin-excitation:
this operator
not only creates a hole on site $j$ but dresses this hole with a 
spin-excitation on a neighboring site, which is exactly the idea behind the 
spin-bag\cite{Schrieffer88,Schrieffer89} or 
spin-polaron\cite{polarons} pictures known in the literature.\\
Splitting this operator into eigenoperators of $H_U$:
\begin{eqnarray}
\hat{C}_{i\uparrow} &=& 
\sum_{j\in N(i)}( \hat{c}_{j,\uparrow} S_i^z + \hat{c}_{j,\downarrow} S_i^-),
\nonumber \\
\hat{D}_{i\uparrow} &=& 
\sum_{j\in N(i)} (\hat{d}_{j,\uparrow} S_i^z + \hat{d}_{j,\downarrow} S_i^-),
\label{eigy}
\end{eqnarray}
we find 
$[\hat{D}_{i\sigma}, H_U] = \frac{U}{2} \hat{D}_{i,j,\sigma}$
and
$[ \hat{C}_{i\sigma}, H_U] = - \frac{U}{2} \hat{C}_{i,j,\sigma}$.
Assuming moreover that the mobility of these composite
excitations is determined by the `heavy' spin-excitation, it seems
quite resonable to assume that these `particles' are the source of the (more or less)
dispersionless bands at $\approx \pm U/2$.\\
Finally, the commutation relation (\ref{commute}) shows that
the mixing matrix element between the $h_{\bbox{k},\sigma}$ and
the new composite particles should be $\approx t$.
Based on these rough considerations we might thus expect that
the two string-$1$ `effective particles' (\ref{eigy}) are excellent candidates
for explaining the two dispersionless bands  at $\pm 3 \, t$ required 
to upgrade the Hubbard-I or SDW approximation so as to match the QMC data.
However, so far the above considerations are pure speculation
and in the following we will turn to QMC-results to back up this hypothesis
by numerical evidence.\\
Before doing so, however, we want to illustrate the action of the `string-operator'
in two extreme cases: an ideal N\'eel state and a resonating valence-bond (RVB) state,
i.e. a compact singlet covering of the plane\cite{Fradkin91} (see figure \ref{fig6}).
In the N\'eel state, 
a hole created initially on site $i$ can travel one place to a neighboring 
site $j$ thereby leaving behind a misaligned spin on the original site $i$.
Exactly this process is described by
the second term, $c^{}_{j,\downarrow} S^{-}_{i}$:
it creates a hole of opposite spin
on a neighboring site $j$ and flips the spin on the original site $i$. 
Therefore, this process corresponds to the creation of a string of length $1$. In fact
one might think
about more sophisticated diagnostic operators incorporating the effects of longer-ranged strings
\cite{Dagotto91}. 
Indeed, Dagotto and Schrieffer\cite{Dagotto91}
and Eder and Ohta\cite{Eder94} 
already measured the angle-resolved spectrum of a diagnostic operator containing 
strings with up to three lattice sites range by means of exact diagonalizations of the $t-J$ model .
As already mentioned in the introduction, in the QMC method each observable has 
to be expressed in terms of free single-particle Greens functions on each time slice by the application 
of Wicks theorem \cite{Fetter71}. This results already in a quite large expression for the spin-$1/2$ 
string operator of equation (\ref{stringoperator}) containing approximately $80$ contributions.
The implementation of even longer-ranged string operators therefore
was not possible.\\
Returning to the spin-$1/2$ string operator of equation (\ref{stringoperator}) we note
that the first term, $c^{}_{i,\uparrow} S^{z}_{j}$ will always annihilate
a N\'eel state. This reflects the simple fact
that spin-rotation symmetry is broken in the N\'eel state.
This is not the case, however, in the fully rotationally invariant RVB
state: again, 
creating a hole on site $i$ and allowing it to hop to site $j$ will produce a
spin-excitation. However, in the case of the RVB state, it produces the superposition
of two states: in one case the dotted ellipse stands for the
$S_z=-1$ component of the triplet, and this state would again be created by the
term $c^{}_{j,\downarrow} S^{-}_{i}$. There is, however, also a second state where the
dotted ellipse corresponds to the superposition of a singlet and the $S_z=0$ component of
the triplet. This second state then would be created by the term $c^{}_{i,\uparrow} S^{z}_{j}$,
whereby the relative sign of the two terms in the string-1 operator makes sure
that the two configurations are always produced with the proper phase.
In both extreme cases, N\'eel state and
`singlet soup', the string-$1$ operator thus creates a hole dressed with the proper
spin-excitation: this scan be a spin wave (i.e. a single inverted spin) 
in the case of a N\'eel state, or
a singlet-triplet excitation in the case of an RVB state.\\
As a technical remark we still note that the excessive
numerical effort which would have been necessary to compute spectra
for the $\hat{C}_{i,\sigma}$ and $\hat{D}_{i,\sigma}$ (which are products of
$5$ Fermion operators) has made it impossible to compute
the spectra of these operators - instead we have been using 
the (Fourier transform of) the 
operator $b^{}_{i,\uparrow}$, defined in  (\ref{stringoperator}).
Concerning the difference between this diagnostic operator
and the operator
$\sum_{j \in N(i)} (c^{}_{j,\uparrow} S^{z}_{i} + c^{}_{j,\downarrow} S^{-}_{i})$ 
obtained by commuting $h^{\dagger}_{i,\uparrow}$ with the kinetic energy
(see the second term on the r.h.s. of equation (\ref{commute})), we note that their 
Fourier transforms 
differ only by phase factors of the form $e^{-i{\bf k}({\bf R}_{i}-{\bf R}_{j})}$. 
Both operators are indeed identical at momentum ${\bf k}=(0,0)$ and differ at momentum 
${\bf k}=(\pi,\pi)$ only by a factor of $-1$ since ${\bf R}_{i}$ and ${\bf R}_{j}$ are 
next-nearest neighbor lattice sites. At all other momenta both operators have exactly the same peaks 
but will differ somewhat in their spectral weights.\\
After these remarks we discuss the angle-resolved spectral function of the 
spin-$1/2$ string operator $b^{}_{{\bf k},\uparrow}$,
shown in figure \ref{fig7}.
The spectrum of the spin-$1/2$ string operator indeed 
picks up those band portions which, according to the rough
hybridization scenario in figure \ref{fig5},
should have strong `flat-band character', i.e. the
part of the narrow low energy band between $(0,0)$ and $(\pi/2,\pi/2)$ and
between  $(0,0)$ and $(\pi,0)$ at energies $\omega \approx -3 \, t$. 
We note that these are precisely the positions where the two quasiparticle-like 
dispersive narrow bands occurred in the normal photoemission spectrum at
$T=0.33 \, t$. In addition, the
band portion at $\omega \approx -6 \, t$ for momentum $(\pi,\pi)$ is also
enhanced in the string-1 spectrum. \\
This, however, still leaves an important part of the band structure unexplained.
Namely, the `AF umklapp band' of the
narrow quasiparticle band dispersing upward between $(0,0)$ and $(\pi/2,\pi/2)$
at $\omega \approx -3 \, t$ still is not seen in any of the spectra, 
not even at the lowest temperature studied.
On the other hand, in a state with true antiferromagnetically broken
symmetry  we know that this mirror image must exist due to the
backfolding of the Brillouin zone.
To finally resolve this part, we now introduce the last diagnostic operator of this work, 
which we call the spin-$3/2$ string operator:
\begin{equation}
\tilde b^{}_{i,\uparrow}=\sum_{j \in N(i)} ({\bf 2} c^{}_{i,\uparrow} S^{z}_{j} - c^{}_{i,\downarrow} 
S^{-}_{j}). \label{string32}
\end{equation}
This describes again a composite object of a hole and a spin-excitation, but this time the
two constituents are coupled to the total spin of $3/2$. We stress that this operator
will detect states which can never be seen in an actual angle-resolved photoemission
spectrum on a singlet ground state, because this is forbidden by the angular-momentum selection rule.
The angle-resolved spectral function $\tilde A({\bf k},\omega)$ of the spin-$3/2$ string operator 
is plotted in figure \ref{fig8} again for $T=0.33 \, t$ (top) and $T=0.10 \, t$ 
(bottom). It is then immediately obvious
that it is this operator which resolves the `missing piece' of the AF dispersion,
i.e. the `AF mirror image' of the narrow quasiparticle band.
It should also be noted that the spectrum of $\tilde b^{}_{i,\uparrow}$ is
remarkably independent of temperature, i.e. the states belonging to this
`spin-$3/2$ band' persist irrespectively of whether there is
long-range order or not.\\
Combining the information obtained so far, 
suggests the following scenario for the crossover between
the paramagnetic band structure at high temperature and the
AF band structure at low temperature:
in the paramagnetic state at high temperatures (such as $T=0.33 \, t$), the spin is a good 
quantum number and the spin-$3/2$ `band' does exist but
cannot mix with any spin-$1/2$ band due to 
spin-conservation.
The band of spin-$3/2$ quasiparticles thus plays no role whatsoever in the
actual photoemission spectrum, which is presumably the reason why the outer part of the
spectrum is so remarkably `invisible' in actual ARPES experiments,
leading to the idea of a `remnant Fermi surface' in the insulator\cite{ronning}.
Reaching this state by photoemission would only be possible
if the photohole is created in a thermally admixed state of at least spin $1$.
In an infinite system this situation changes 
discontinuously at the transition to the true broken symmetry state:
there the total spin ceases to be a good quantum number, and the
spin-$1/2$ band in the interior of the AF zone
and the spin-$3/2$ band in the exterior now suddenly can mix with each other,
thus leading to the familiar SDW dispersion.
We note that another way to generate a coupling between these
two bands would be application of a magnetic field - this also would break 
spin-rotation invariance and hence enable the hybridization of
a spin-$1/2$ and a spin-$3/2$ band. Based on our results we thus believe that
a magnetic field could enhance the spectral weight of the
`shadow part' of the band structure as seen in ARPES.
\section{Doping the Hubbard model at high temperatures -- rigid bands}
Summarizing our results so far, we may say that the Hubbard-I approximation, slightly
improved by the introduction of new quasiparticles corresponding to dressed holes
provides a very good description of the spectral function for the case
of half-filling, $\langle n \rangle = 1$.
In this section we want to proceed to the doped case $\langle n \rangle < 1$, 
which is of prime interest for
cuprate superconductors. Here, an essential drawback of the QMC procedure
is that reliable QMC 
simulations for lower temperatures are much more difficult or even impossible, since the 
absence of particle-hole symmetry away from half-filling introduces the notorious minus-sign
problem into the algorithm. 
Truely low temperatures like $T=0.10 \,t$, which in principle
correspond to the physical temperature range, are therefore out of reach.
On the other hand in the study of the half-filled case
we have seen that a major change takes place as the spin correlation length
reaches the system size, whence `effective long range order'
sets in. In the doped case the spin correlation length is expected to
be short at any temperature, whence we may expect that the
change of $A({\bf k},\omega)$ from high to small temperature is more
smooth than at half-filling. In that sense, even $A({\bf k},\omega)$
data for the relatively
high temperature $T=0.33 \, t$ are interesting to study.
Moreover, we can at least try to elucidate trends with
decreasing temperature and thus construct a reasonably
plausible scenario.\\
At half-filling, we have seen that the `approximation of choice' for the
paramagnetic case was the Hubbard-I approximation.
This naturally poses the question as to how
relevant the half-filled case is for the description of the doped case, i.e.
how much of the Hubbard-I physics remains valid for finite doping.
At half-filling the two `effective particles' $\hat{d}_{i,\sigma}^\dagger$ and
$\hat{c}_{i,\sigma}^\dagger$, form the two separate Hubbard bands.
The effect of doping would now consist in the chemical potential cutting
progressively into the top of the lower Hubbard band, in much the same
fashion as in a doped band insulator. On the other hand, for
finite $U/t$ the spectral weight
along this band deviates from the
free-particle value of $1$ per momentum and spin so that the Fermi surface
volume (obtained from the requirement that the integrated
spectral weight up to the Fermi energy be equal to the total number 
of electrons) is not in any `simple' relationship to the
number of electrons - the Luttinger theorem must be violated. This is the major
reason why the Hubbard-I approximation has been dismissed by many authors as
being unphysical.\\
We now wish to address the question as to what really
happens if a paramagnetic (i.e. not magnetically ordered) 
insulator is doped away from half-filling, by
QMC simulation. We therefore choose $T=0.33t$ and $U=8.0 \, t$.\\
Figure \ref{fig9} then shows the development of $A(\bbox{k},\omega)$
with doping. 
It is quite obvious from this Figure that initially the $2$ bands seen at half-filling
in the photoemission spectrum (i.e. $\omega < 0$) persist with an essentially unchanged
dispersion. The chemical potential gradually cuts deeper and deeper
into the topmost band, forming a hole-like Fermi surface centered
on $(\pi,\pi)$, the top of the lower Hubbard band.
The only deviation from a rather simple rigid-band behavior is
an additional transfer of spectral weight:
the part of the topmost band near $(\pi,\pi)$ gains in spectral weight,
whereas the band with higher binding energy looses weight. 
In addition, there is a transfer of weight
from the upper Hubbard band to the inverse
photoemission part below the Hubbard gap.
This effect is actually quite well understood\cite{henk}.
The band structure above the Hubbard gap
becomes more diffuse upon
hole doping in that the rather clear
two-band structure visible near $(\pi,\pi)$ at half-filling
rapidly gives way to one broad `hump' of weight.
Apart from the spectral weight transfer, however, the
band structure on the photoemission side is almost unaffected
by the hole doping - the {\em dispersion} of the
quasiparticle band becomes somewhat wider but does not change appreciably. 
In that sense we see at least qualitatively the behavior
predicted by the Hubbard I approximation.\\
Next, we focus on the Fermi surface volume. Some care
is necessary here: first,
we cannot actually be sure that at the high temperature we are using there
is still a well-defined Fermi surface. Second, the criterion we will be using
is the crossing of the quasiparticle band
through the chemical potential. It has to be
kept in mind that this may be quite misleading, because band portions with
tiny spectral weight are ignored in this approach (see for example
Ref. \cite{comment} for a discussion).
When thinking of a Fermi surface as the constant energy contour of the
chemical potential, we have to keep in mind that
portions with low spectral weight may be overlooked.
On the other 
hand the fact that a peak with appreciable weight crosses 
from photoemission to inverse photoemission
at a certain momentum is independent of whether we call this a
`Fermi surface' in the usual sense, and should be reproduced
by any theory which claims to describe the system.
It therefore has to be kept in mind that in the following
we are basically studying a `spectral weight Fermi surface', i.e.
the locus in ${\bf k}$ space where an apparent quasiparticle band
with high spectral weight crosses the chemical potential.
With these {\em caveats } in mind, Figures \ref{fig10} and
\ref{fig11} show the low-energy
peak structure of $A(\bbox{k},\omega)$ for all allowed momenta
of the $8\times 8$ cluster in the
irreducible wedge of the Brillouin zone, and 
for different hole concentrations.
In all of these spectra there is a pronounced peak,
whose position shows a smooth dispersion with momentum.
Around $(\pi,\pi)$ the peak is  above $\mu$, whereas in the
center of the Brillouin zone it is below.
The locus in $\bbox{k}$-space where the peak crosses $\mu$ forms
a closed curve around $(\pi,\pi)$ and it is obvious
from the Figure that the `hole pocket' around
$(\pi,\pi)$ increases very rapidly with
$\delta$. To estimate the Fermi surface volume $V_F$
we assign a weight $w_{\bf k}$ of $1$ to momenta ${\bf k}$ where the
peak is below $\mu$, $0.5$ if the peak is right at $\mu$ and
$0$ if the peak is above $\mu$. Our assignments of these weights
are given in Figures \ref{fig10} and \ref{fig11}. 
The fractional Fermi surface volume then is
$V_F = \frac{1}{N}\sum_{\bf k} w_{\bf k}$, where $N=64$ is the
number of momenta in the $8\times 8$ cluster.
Of course, the assignment of the $w_{\bf k}$
involves a certain degree of arbitrariness.
It can be seen from Figures
\ref{fig10} and \ref{fig11}, however, that our  $w_{\bf k}$ 
would in any way tend to underestimate the Fermi surface volume, so 
that the obtained $V_F$ data points rather have the character of a 
lower bound to the true $V_F$. Even if we take into account some
small variations of $V_F$ due to different assignments
of the weight factors, however, the resulting $V_F$ versus $\delta$ 
curve never can be made consistent with the Luttinger volume, see
Figure \ref{fig12}. The deviation from the Luttinger 
volume is quite pronounced at low doping.
$V_F$ approaches the Luttinger volume for dopings
$\approx 20$\%, but due to our somewhat crude way of
determining $V_F$ we cannot really decide when precisely 
the Luttinger theorem is obeyed. The Hubbard I approximation
approaches the Luttinger volume for hole concentrations of $\approx 50$\%,
i.e. the steepness of the drop of $V_F$ is not reproduced
quantitatively. The latter is somewhat improved in 
the so-called $2$-pole approximation\cite{bernhard,beenen}. For
example the Fermi surface given by
Beenen and Edwards\cite{beenen} for $\langle n \rangle=0.94$ obviously
is very consistent with the spectrum in Figure \ref{fig11} for
$ \langle n \rangle=0.95$.\\
We return to Figure \ref{fig9} and
discuss the entire width of the spectra, in particular the
question of the fate of the 4-band structure in the doped system.
For $\langle n \rangle=0.95$
the different features that are seen at $\langle n \rangle=1.0$
are still rather clearly visible, but for $\langle n \rangle=0.86$ the low energy
quasiparticle band at $k=(0,0)$ starts to disappear, and at $\langle n \rangle=0.80$
the dominant `band' in the spectrum between $\omega=-4 \, t$ and
$\omega=2 \, t$ can be fitted by a sligtly renormalized free-electron
band. As we have seen above, the Luttinger theorem also is
valid in this case. This suggests to classify the doping as
`underdoped' for $0<\langle n \rangle<0.85$, where the Luttinger theorem is invalid and the
$4$-band structure known from half-filling persists,
and `overdoped' where the Luttinger theorem is valid and
a renormalized free-electron band can be seen in the spectral function.
Following the convention for cuprate superconductors, we
call the doping where the crossover between the two regimes
occurs the `optimal' doping.\\
Next, the four plots of figure \ref{fig13} show the spectra
at selected $\bbox{k}$-points.
The system size is only $6 \times 6$ in this case because this allows for
smaller error bars. Closer inspection, especially, of the peaks at 
momentum ${\bf k}=(0,0)$ on the photoemission side (plot (a)) and at momentum ${\bf k}=(\pi,\pi)$
on the inverse photoemission side (plot (d)) confirms, that with increasing hole
concentration we are losing parts of the 4-band structure seen at half-filling.\\
To check the physics of the band structur in more detail, we again employ our diagnostic operators.
Figure \ref{fig14} shows the angle-resolved spectral functions 
$\tilde A({\bf k},\omega)$ of the spin-$1/2$ and spin-$3/2$ string operators, $b^{}_{i,\uparrow}$ and
$\tilde b^{}_{i,\uparrow}$. As was the case at half-filling,
the spectrum of the spin-$1/2$ string operator 
highlights exactly those peaks that we associate with
the dipsersionless `dressed hole' bands in Figure \ref{fig5}.
The spectrum of the spin-$3/2$ string operator on the other hand has its
peaks with maximal spectral weight around momentum ${\bf k}=(\pi,\pi)$,
indicating that also in the doped case there is an `antiferromagnetic mirror image'
of the quasiparticle band (which, however, consists of spin-$3/2$ states).
Again, coupling of photoholes to thermally excited spin excitations may make these
states visible in ARPES spectra, thus explaining the `shadow bands' seen in
photoemission experiments by Aebi {\em et al.}\cite{Aebi}.
Similarly as for half-filling one might speculate that a magnetic field, which
would break spin symmetry and thus allow for a coupling of
`bands' with different total spin, would enhance the spectral weight
of these shadow bands.\\
All in all we have ssen that the `band structure' ($4$-band structure, dispersion of regions 
of large spectral weight, `character' of the bands as measured by the diagnostic operators)
stays pretty much unchanged as long as we are in the underdoped regime.
At half-filling the $4$-band structure is closely related to
the sharp low-energy mode in the dynamical spin correlation function, which
naturally suggests to study the spin reponse also as a function of doping.
Figure \ref{fig15} shows the spin-correlation function, $\chi_{sz}
({\bf k},\omega)$ (left column), and the charge-correlation function, $\chi_{cc}({\bf k},\omega)$
(right column), for  $T=0.33 \, t$ and densities $\langle n \rangle=0.95$ (underdoped), 
$\langle n \rangle=0.90$ (nearly optimally doped) and $\langle n \rangle=0.80$ (overdoped). 
The spin-response is sharply confined in both 
momentum ${\bf k}=(\pi,\pi)$ and energy $\omega=\omega^{\star}$ only in the underdoped region 
i.e. the regime where we also observe the features associated with
spin excitations in the single-particle spectra.
As was the case at half-filling for temperatures below $T \approx 0.33 \, t$, 
the spin-response can be fitted by the AF spin-wave dispersion (\ref{spinwave}) in
the underdoped regime. On the 
other hand, as soon as the system enters the overdoped regime the spin-response is no longer 
sharply peaked at momentum ${\bf k}=(\pi,\pi)$ and energy $\omega=\omega^{\star}$: it broadens 
in momentum and spreads in energy by an order of magnitude with the scale changing from 
$J=4t^{2}/U$ to $E_{kin} \sim 8.0 \, t$ accompanied by a similar change in the bandwidth 
of the single particle excitations. This result is already well known from previous QMC 
calculations \cite{Preuss97} and consistent with similar behaviour in the
t-J model\cite{jaklic1}.
The charge-response, $\chi_{cc}({\bf k},\omega)$, is always broad in both momentum ${\bf k}$ and 
energy $\omega$ for all densities studied. It merely decreases its width from $\approx 12.0 \, t$ at 
$\langle n \rangle=0.95$ to $\approx 8.0 \, t$ at $\langle n \rangle=0.80$.\\
Although the minus-sign problem of the QMC algorithm prevents reliable simulations of 
large systems at low temperatures in the doped regime, we nevertheless studied the temperature 
evolution of the angle-resolved spectral function $A({\bf k},\omega)$ at density $\langle n 
\rangle=0.93$. This was possible due to the relative small system size of $6 \times 6$, which 
alleviates the minus-sign problem as compared to $8 \times 8$ at the  $T=0.25 \, 
t$. Figure \ref{fig16} shows the results from this analysis: the
uppermost plot (a) compares the angle-resolved spectral functions $A({\bf k},\omega)$ at density
$\langle n \rangle=0.93$ for $T=0.50 \, t$, $T=0.33 \, t$ and $T=0.25 \, t$.
We stress that the simulation at  $T=0.25 \, t$ suffers from minus-sign problems
with a drastically reduced resolution. In the center plot (b), the quasiparticle peak weights 
around momentum ${\bf k}=(\pi,\pi)$ of the $\langle n \rangle=0.93$ simulation are compared 
with the quasiparticle peak weight at momentum ${\bf k}=(\pi,\pi)$ of a half-filled, $\langle
n \rangle=1.0$, simulation for different temperatures. 
At half-filling, the Hubbard-I-like quasiparticle peak 
${\bf k}=(\pi,\pi)$ and $\omega \approx -1.5 \, t$ 
decreases in spectral weight with decreasing $T$ and disappears as the 
spin-correlation length (which increases with decreasing temperature) reaches the lattice size 
(at $T \approx 0.20 \, t$). In the underdoped case for density $\langle n \rangle=0.93$ 
the weights of the corresponding
peaks around momentum ${\bf k}=(\pi,\pi)$ located
also decrease with decreasing $T$. Closer inspection 
of this peak (see the inset of the center plot (b)) reveals, that this peak even raises slightly 
in binding energy with decreasing temperature, very similar to the peak in the half-filled case. 
In a real photoemission experiment this peak would have dropped below the typical 
resolution of roughly $10 \%$ spectral weight \cite{Haas95} at a temperature of $T \approx 0.25 
\, t$. The spin-correlation length (again derived by a fit of the equal-imaginary-times 
spin-correlation function to a form $a \cdot \vert {\bf r} \vert^{b} \cdot e^{-\vert {\bf r} 
\vert/\zeta_{sz} (T)}$) also shows similar behavior in the underdoped and half-filled 
cases: the values for the spin-correlation length are $\zeta_{sz}=0.5-0.8$, $\zeta_{sz}=0.8-1.0$ 
and $\zeta_{sz}=1.0-1.3$ in the underdoped case and $\zeta_{sz}=0.6-0.9$, $\zeta_{sz}=1.0-1.3$ and 
$\zeta_{sz}=1.6-1.9$ at half-filling for $T=0.50 \, t$, $T=0.33 \, t$ and 
$T=0.25 \, t$, respectively. The spin-susceptibility (shown in plot (c) of figure 
\ref{fig16}) also behaves very similar, but with changed magnitudes in the underdoped and in 
the half-filled cases. These data suggest a similar temperature evolution of the band-structure 
of the Hubbard model in the underdoped and half-filled cases driven by the temperature dependent 
spin-correlation length $\zeta_{sz} (T)$. Especially, we expect the Hubbard-I-like quasiparticle peaks 
at energies of $\omega \approx 1.0 \, t$ around momentum ${\bf k}=(\pi,\pi)$ to vanish with decreasing 
 $T$ in the underdoped case as the peaks at energies of $\omega \approx -1.5 \, t$ around 
momentum ${\bf k}=(\pi,\pi)$ do in the case of half-filling.\\
The latter observation suggests a profound change of the Fermi surface with
temperature: as seen above, it is precisely
the Hubbard-I-like band near $(\pi,\pi)$ which crosses the chemical potential and
thus forms the Fermi surface in the doped case. It is then quite clear that the
`disappearance' of this band with decreasing temperature must affect the Fermi surface
in some dramatic way. 
Studies at zero temperature are possible only by means of 
exact diagonalization. Analysis of the single particle spectrum shows
the same `rigid-band' behaviour as at high temperatures\cite{shimozato} and analysis of
the momentum distribution $n(\bbox{k})$ suggest\cite{nishimoto} that the
doped holes accumulate at the surface of the magnetic zone (i.e. the line
$(\frac{\pi}{2},\frac{\pi}{2})\rightarrow (\pi,0)$) rather than
around $(\pi,\pi)$.
\section{Summary}
In the present work we have systemactically studied 
the temperature- and doping-dependent dynamics of the
two-dimensional Hubbard model by finite-temperature QMC simulations. 
Comparing the QMC single particle spectral function,
the dynamical spin response and
the spectral functions of suitably chosen diagnostic operators, different physical
regimes could be identified. In simplest terms there are two quantities, which basically determine the
single particle spectrum: the hole concentration and the
the spin-response function, whereby there is a certain relationship between the two.\\
At half-filling and
high temeratures ($T \geq t$), the combined
photoemission and and inverse photoemission spectrum
$A({\bf k},\omega)$ displays two dispersive features, the upper and
lower Hubbard band, roughly separated by $U$ ($=8 \, t$, in our
work). At these very high temperatures the system is in a spin disordered state. We have
demonstrated here that the well-known Hubbard-I approximation gives an excellent description
of the single particle spectrum
in this state, reproducing quantitatively both the single-particle dispersion
and the distribution of spectral weight. This is by no means trivial, since the 
Hubbard-I approximation is dynamically equivalent to a simplified effective Hamiltonian,
which just contains hole-like ($h^{\dagger}_{i,\sigma}$) and double occupancy-like 
particles in a simple bi-quadratic form.\\
At lower temperatures ($T \leq 0.33 \, t$), the Hubbard-I approximation needs to be improved;
this is to be expected, because it neglects all effects of spin-correlations. In fact, the
temperature where deviations from Hubbard-I become strong, coincides fairly well with
the transition from a spin response $\chi_{sz}({\bf k},\omega)$
which is diffuse both in momentum and energy (with a spread of order $t$), to a more
`spin-wave-like' response. In this regime
$\chi_{sz}({\bf k},\omega)$ displays the characteristic energy scale $J=4t^2/U$, with its 
spectral weight being concentrated at the AF wave vector ${\bf Q}=(\pi,\pi)$. 
It should be noted that this `spin-wave-like' regime develops
despite the fact that at $T=0.33 \, t$ the spin-correlations length $\zeta_{sz} 
(T)$ is still short-ranged ($\leq 2$ lattice spacings). Only at the
lower temperature $T=0.10 \, t$ N\'eel order spreads over the entire QMC block,
creating an effective (finite-size) N\'eel state.\\
It is well established by previous, in particular also QMC work, that in this
temperature-regime ($T \leq 0.33 \, t$) new spectral features appear. They 
have often been interpreted as four `bands', two `coherent' bands forming
the topmost valence and the lowest conduction band in the insulator plus
two `incoherent' bands, i.e. the remaining upper and lower Hubbard band 
features (see for example \cite{Preuss95}). Our present work not only 
definitively identifies these four bands but also clarifies their physical
origin and their connection to the spin-excitations. 
In simplest terms the emerging spin waves at lower
temperatures provide the excitations that can
`dress' the  Hubbard quasiparticles, whence new bands
corresponding to dressed holes/double occupancies appear in the single particle spectrum
$A({\bf k},\omega)$.
It has been shown in Ref. \cite{ansgar} that
the $4$-band structure which appears in $A({\bf k},\omega)$ at lower temperatures
can be explained in this way,
and our present numerical check by directly calculating the spectra of `dressed
electrons' supports this interpretation.\\
This physical picture at half-filling can be extended into the underdoped
regime. This is most obvious in the single particle spectral function,
which stays almost unchanged in the doped case
(i.e. the $4$-band structure and the `character' of the bands as measured by the
diagnostic operators). The main change in fact consists in the
chemical potential cutting gradually into the (top of the) lower Hubbard band,
precisely as predicted by the Hubbard-I approximation. Contrary to widespread
belief the `Fermi surface', if determined by the Fermi surface crossings of the
dominant band through the chemical potential, does not satisfy the Luttinger theorem.
Rather, for small hole concentrations the Fermi surface volume is considerable larger
than that for a slightly less than half-filled free-electron band.
Very similar conclusions have in fact been reached by
a calculation of the electron momentum distribution in the 2D t-J model
by Puttika {\em et al.}\cite{Puttika}. Their calculation actually was a
high temperature series expansion plus a Pad\'e extrapolation to lower temperature, and it
is encouraging that this method gives similar results as our
QMC results which are performed at relatively high temperatures. 
In its range of applicability, i.e. in the absence of
strong magnetic correlations and close to half-filling, 
the Hubbard-I approximation thus works remarkably well,
both at half-filling and in the doped case. We stress that this has profound
implications for the theoretical treatment of the model:
perturbation expansions in $U$ or partial and self-consistent resummations
thereof, may not be expected to give any meaningful results in this
strong-coupling/low doping regime.\\
An interesting question is the possibility to verify our results
experimentally. As already mentioned above, a scan of the temperature development of
$A(\bbox{k},\omega)$ shows that the part of the quasiparticle band near $(\pi,\pi)$
(where the Fermi surface is located) is loosing weight with decreasing
temperature. In fact in ARPES experiments on underdoped cuprate superconductors the
`hole-pocket' around $(\pi,\pi)$ seen in our simulations
(and the expansion of Puttika {\em et al.}) is not observed, but rather
a small `Fermi arc' near $(\pi/2,\pi/2)$, terminated by the
`pseudo gap' around $(\pi,0)$
Although our simulations do not allow to make statements about the
truely low temperatures in the experiments, we believe that this
suggests a strong temperature dependence of the single particle spectrum, with the
temperature scale being set by the exchange constant $J$ (which controls the degree of
spin disorder). The latter is rather large in copper oxides, so that
the temperature regime studied in our simulations probably cannot be accessed
experimentally in these materials. We note, however, that an ARPES study for the 1D material
$Na_{0.96}V_{2}O_{5}$
which has a smaller exchange constant, has indeed provided evidence for a strong
$T$-dependence of $A(\bbox{k},\omega)$\cite{Kobayashi}. Clearly, it would be interesting to
study the Fermi surface evolution in a 2D material with lower
exchange constant.\\
As was the case at half-filling, the dynamical spin-response plays an important role:
throughout the Hubbard-I phase at low doping, the spin response shows the
sharp low-energy mode at $(\pi,\pi)$. The simultaneous
disappearance of the $4$-band structure in $A(\bbox{k},\omega)$ and the
low energy spin response with scale $J$ in the overdoped regime then show
again the close relationship between the two. The dressing of holes by
spin excitations apparently remains the most important correction over Hubbard-I.
In the overdoped regime the
spin response is spread out over an energy range of $\approx 8t$ and thus becomes
more similar to the charge response. The
single-particle spectral function is most consistent with
a slightly renormalized free electron dispersion, and the Luttinger theorem appears
to be satisfied even at the relatively high temperature $T=0.33t$.
This is quite consistent with earlier results on the t-J model, which show
that for hole concentrations $\ge 25$ \% the spin and charge response can be
approximated well by the self-convolution of the single particle spectral function\cite{highdoping}.
This is essentially what is to be expected for a system of weakly interacting Fermions,
so that we conclude that in the overdoped regime we enter a new
phase which most probably extends to the low-concentration limit where the Nagaoka $T$-matrix
approximation becomes exact. Finally we note that exact diagonalization studies at
finite temperatures\cite{jaklic} also show some evidence for a
`crossover' between different physical regimes at a hole concentration of
$15$\%.\\
This work was supported by DFN Contract No. TK 598-VA/D03 and by BMBF (05SB8WWA1). 
Computations were performed at HLRS Stuttgart and HLRZ J\"ulich. One of us (W.H.)
acknowledges hospitality of the Physics Department in Santa Barbara and many useful discussions
with D. J. Scalapino.

\end{multicols}

\newpage

\begin{figure}[htbp!]
  \vspace{0.0cm}\hspace{0.00cm}
  \vspace{0.0cm}\hspace{0.00cm}\widetext
  \caption[]{Angle-resolved spectral function $A({\bf k},\omega)$ of a $8 \times 8$ Hubbard lattice 
    at $\langle n \rangle = 1.0$ and $U=8.0 \, t$ for various temperatures.
    The solid lines in the left column represent the
    (renormalized) Hubbard-I and SDW dispersions. In the right column the spectral functions from 
    QMC (solid lines) are compared to the Hubbard-I and SDW-approximations with a Lorentzian 
    lineshape (dashed lines).}
  \label{fig1} 
\end{figure}
\begin{figure}[htbp!]
  \vspace{0.0cm}\hspace{0.00cm}
  \vspace{0.0cm}\hspace{0.00cm}\widetext
  \caption[]{Dynamical spin-correlation 
    function, $\chi_{sz}({\bf k},\omega)$ (left column), and charge-correlation 
    functions $\chi_{cc} ({\bf k},\omega)$ (right column), of a $8 \times 8$ Hubbard lattice 
    at $\langle n \rangle = 1.0$ and $U=8.0 \, t$ for $T=1.00 \, t$, $T=0.33 \, t$ 
    and $T=0.10 \, t$. The solid lines in the lower left column give a spin-wave 
    fit.}
  \label{fig2} 
\end{figure}
\begin{figure}[htbp!]
  \vspace{0.0cm}\hspace{0.00cm}
  \vspace{0.0cm}\hspace{0.00cm}\narrowtext
  \caption[]{Angle-resolved spectral function $\tilde A({\bf k},\omega)$ of the shadow-operator
    $\tilde c^{}_{i,\sigma}$
    with $8 \times 8$ lattice size at $\langle n \rangle = 1.0$ and $U=8.0 \, t$ for s
    $T=0.33 \, t$ and $T=0.10 \, t$.}
  \label{fig3} 
\end{figure}
\begin{figure}[htbp!]
  \vspace{0.0cm}\hspace{0.00cm}
  \vspace{0.0cm}\hspace{0.00cm}\narrowtext
  \caption[]{Single-particle spectral function of the normal photoemission (top) and the shadow-operator 
    (bottom) at momentum ${\bf k}=(\pi,\pi)$ with $8 \times 8$ lattice size at $\langle n \rangle = 1.0$ 
    and $U=8.0 \, t$ for temperatures in between $T=1.00 \, t$ and $T=0.10 \, t$.}
  \label{fig4} 
\end{figure}
\begin{figure}[htbp!]
  \vspace{0.0cm}\hspace{0.00cm}
  \vspace{0.0cm}\hspace{0.00cm}\narrowtext
  \caption[]{Angle-resolved spectral function $A({\bf k},\omega)$ of the $8 \times 8$ Hubbard
    lattice with $U=8.0 \, t$ for  $T=0.33 \, t$ (top) and  $T=0.10 \, 
    t$ (bottom) compared to analytic dispersions: mixing the Hubbard-I (top, dashed) or 
    SDW (bottom, dashed) bands with
    two dispersionless bands at $\omega = \pm 3 \, t$ (dashed) results in a 4-band
    structure (solid) in good agreement with the QMC peaks.}
  \label{fig5}
\end{figure}
\begin{figure}[htbp!]
  \vspace{0.0cm}\hspace{0.00cm}
  \caption[]{Motion of an aded hole in an N\'eel state (left) and an RVB state (right)
   produces spin defects of different kind. These are described by the spin-1/2 string
   operator.}
  \label{fig6}
\end{figure}

\begin{figure}[htbp!]
  \vspace{0.0cm}\hspace{0.00cm}
  \vspace{0.0cm}\hspace{0.00cm}\narrowtext
  \caption[]{Angle-resolved spectral function $\tilde A({\bf k},\omega)$ of the 
    spin-$1/2$ string operator $b^{}_{i,\uparrow}$
    in the Hubbard model with $U=8.0 \, t$ 
    for  $T=0.33 \, t$ (top) and  $T=0.10 \, t$ (bottom). The lattice 
    size was reduced to $6 \times 6$ in this case due to the larger error bars of the 
    string operator as compared to the normal photoemission spectrum.}
  \label{fig7}
\end{figure}
\begin{figure}[htbp!]
  \vspace{0.0cm}\hspace{0.00cm}
  \vspace{0.0cm}\hspace{0.00cm}\narrowtext
  \caption[]{Angle-resolved spectral function $\tilde A({\bf k},\omega)$ of the 
    spin-$3/2$ string operator $\tilde b^{}_{i,\uparrow}$
    in the Hubbard 
    model with $U=8.0 \, t$ for  $T=0.33 \, t$ (top) and  
    $T=0.10 \, t$ (bottom). The lattice size was reduced to $6 \times 6$ in this case 
    due to the larger error bars of the quartet operator as compared to the normal
    photoemission spectrum.}
  \label{fig8}
\end{figure}
\begin{figure}[htbp!]
  \vspace{0.00cm}\hspace{0.00cm}
  \vspace{0.00cm}\hspace{0.00cm}\widetext
  \caption[]{Overview on the doping dependence of the angle-resolved spectral function 
    $A({\bf k},\omega)$ of the $8 \times 8$ Hubbard lattice at $T=0.33 \, t$ and $U=8.0 
    \, t$ for densities $\langle n \rangle=1.00$ (half-filled), $\langle n \rangle=0.95$
    (underdoped), $\langle n \rangle=0.86$ (roughly optimal doped) and $\langle n \rangle=
    0.80$ (overdoped).}
  \label{fig9}
\end{figure}
\begin{figure}
\epsfxsize=6.0cm
\vspace{-0.5cm}
\hspace{1.0cm}
\vspace{0.25cm}
\narrowtext
\caption[]{Single particle spectral function for all
${\bf k}$-points of the $8\times 8$ cluster in the irreducible
wedge of the Brillouin zone. For each ${\bf k}$ the weight
$w_{\bf k}$ is given.}
\label{fig10} 
\end{figure}
\begin{figure}
\epsfxsize=6.0cm
\vspace{-0.5cm}
\hspace{1.0cm}
\vspace{0.25cm}
\narrowtext
\caption[]{Same as Figure \ref{fig10} for lower electron densities.}
\label{fig11} 
\end{figure}
\begin{figure}
\epsfxsize=6.0cm
\vspace{-0.5cm}
\hspace{1.5cm}
\vspace{-0.0cm}
\narrowtext
\caption[]{Fermi surface volume as estimated from the
single particle spectral function, plotted versus
the concentration of holes in the half-filled band.
The dashed line gives the value predicted by the Luttinger theorem,
$V_F=\frac{n}{2}$.}
\label{fig12} 
\end{figure}
\begin{figure}[htbp!]
  \vspace{0.0cm}\hspace{0.00cm}
  \vspace{0.0cm}\hspace{0.00cm}\widetext
  \caption[]{Angle-resolved spectral function $A({\bf k},\omega)$ of a $6 \times 6$ Hubbard lattice 
    at $T=0.33 \, t$ and $U=8.0 \, t$ for momenta ${\bf k}=(0,0)$ (plot (a)) and ${\bf k}=(\pi,0)$ 
    (plot (b)) and ${\bf k}=(\pi,\pi)$ (plot (c) and (d)). The spectra for the various densities 
    $\langle n \rangle=1.00$, $\langle n \rangle=0.96$, $\langle n \rangle=0.90$, $\langle n 
    \rangle=0.83$ and $\langle n \rangle=0.66$ are shifted by their individual chemical potentials 
    $\mu$ resulting in overlapping peaks in the underdoped regime. The labels at the $\omega$-axis 
    refer to the half-filled simulation (solid line). The system size is $6 \times 6$, 
    because the smaller errors in this case lead to more reliable spectra.}
  \label{fig13} 
\end{figure}
\begin{figure}[htbp!]
  \vspace{0.0cm}\hspace{0.0cm}
  \vspace{0.0cm}\hspace{0.0cm}\widetext
  \caption[]{Angle-resolved spectral function $\tilde A({\bf k},\omega)$ of the spin-$1/2$ 
    string operator $b^{}_{i,\uparrow}$ (top) and the spin-$3/2$ string operator 
    $\tilde b^{}_{i,\uparrow}$ (bottom) in the Hubbard model with $U=8.0 \, t$ for 
     $T=0.33 \, t$ and densities $\langle n \rangle=0.95$ (left) and 
    $\langle n \rangle=0.90$ (right).  The lattice size was $6 \times 6$ in this case 
    due to the larger error bars of the string operators as compared to the normal 
    photoemission spectrum.}
  \label{fig14}
\end{figure}
\begin{figure}[htbp!]
  \vspace{0.00cm}\hspace{0.00cm}
  \vspace{0.00cm}\hspace{0.00cm}\widetext
  \caption[]{Dynamical spin-, $\chi_{sz}({\bf k},\omega)$ (left column), and charge-correlation 
    functions, $\chi_{cc}({\bf k},\omega)$ (right column), of a $8 \times 8$ Hubbard lattice 
    at $U=8.0 \, t$ and  $T=0.33 \, t$ for densities $\langle n \rangle=0.95$, 
    $\langle n \rangle=0.90$ and $\langle n \rangle=0.80$. The solid lines in the upper 
    plots of the left column give a spin-wave fit.}
  \label{fig15}
\end{figure}
\begin{figure}[htbp!]
  \vspace{0.0cm}\hspace{0.0cm}
  \vspace{0.0cm}\hspace{-0.15cm}
  \vspace{0.0cm}\hspace{0.0cm}\narrowtext
  \caption[]{(a) Angle-resolved spectral function $A({\bf k},\omega)$ of the $6 \times 6$ Hubbard 
    lattice with $U=8.0 \, t$ and density $\langle n \rangle=0.93$ versus  $T$, (b)
    quasiparticle peak weight around momentum ${\bf k}=(\pi,\pi)$ for $\langle n \rangle=1.0$ 
    ($8 \times 8$) and $\langle n \rangle=0.93$ ($6 \times 6$) versus  $T$ and (c)
    magnetic susceptibility $\chi_{sz}({\bf k}=(\pi,\pi),\omega=0)$ for densities $\langle n 
    \rangle=1.0$ ($8 \times 8$) and $\langle n \rangle=0.93$ ($6 \times 6$) versus  $T$.
    Please note the slightly raising (in binding energy) and vanishing (in spectral weight) with 
    decreasing temperature of the peak located at ${\bf k}=(\pi,\pi)$ magnified in the inset of 
    plot (b). The system size is only $6 \times 6$ in the doped case to avoid the serious sign 
    problems of the $8 \times 8$ system at $T=0.25 \, t$. But also the $6 \times 6$ data at 
    $T=0.25 \, t$ suffer from sign problems at some momenta, ${\bf k}=(0,0)$ and ${\bf k}=
    (\pi/3,0)$.}
  \label{fig16}
\end{figure}
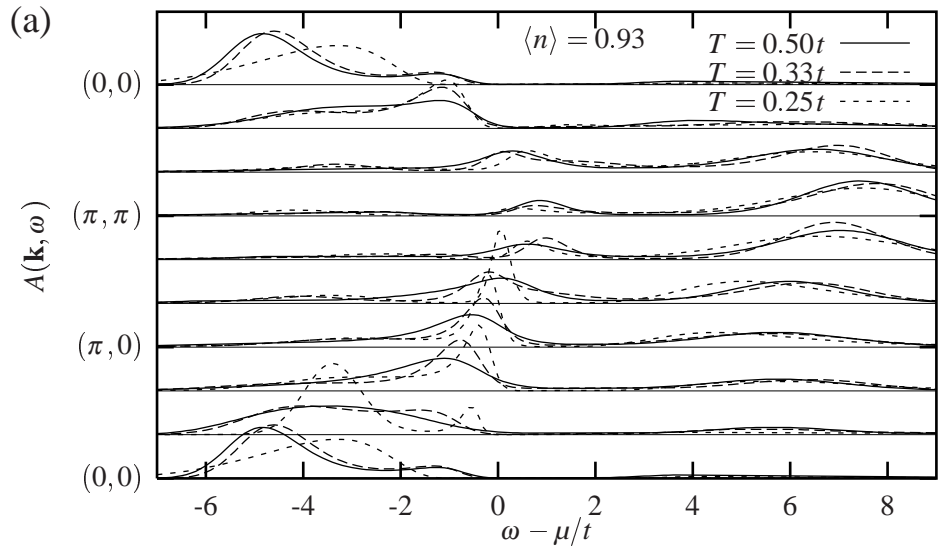
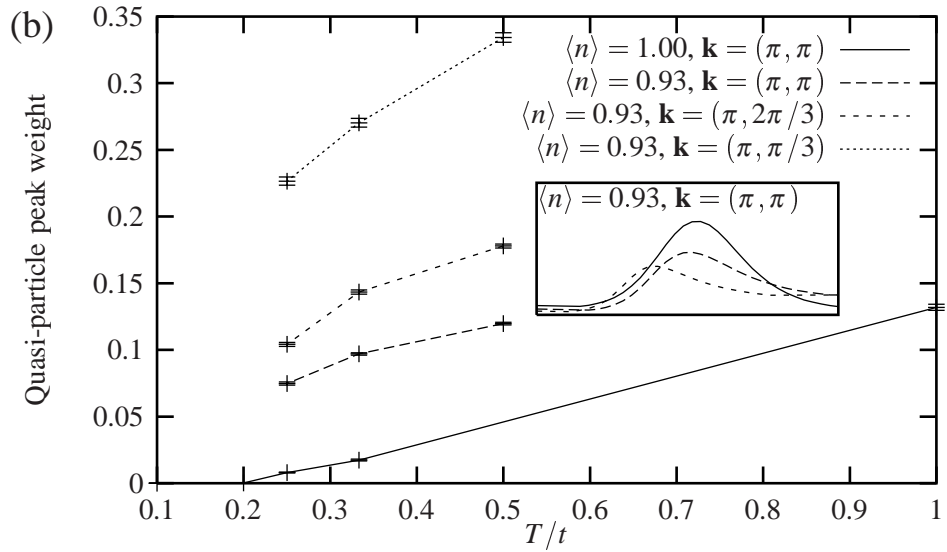
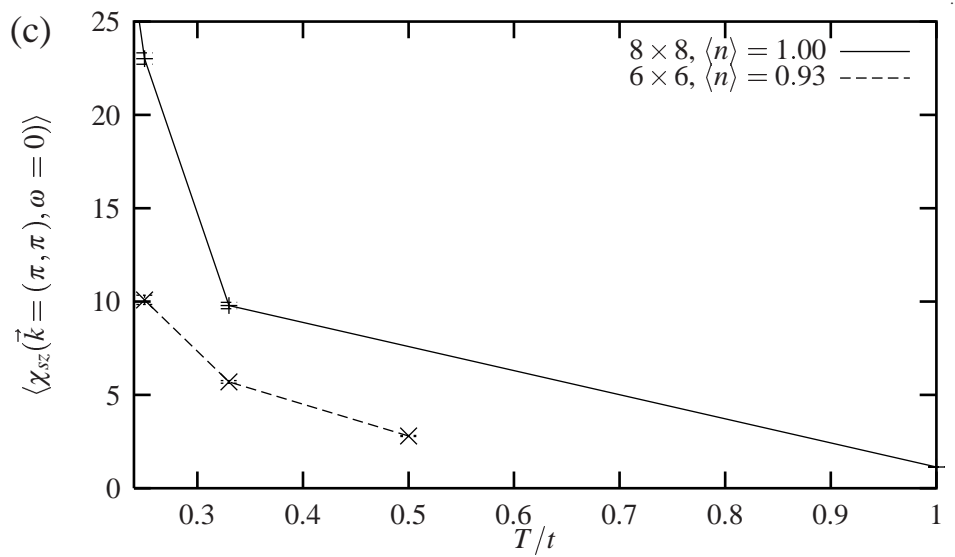
\end{document}